\documentclass[aps,pre,twocolumn,amsmath,amssymb,notitlepage,noeprint,nofootinbib,superscriptaddress]{revtex4-1}
\pdfoutput=1
\usepackage[latin9]{inputenc}
\usepackage{braket}
\usepackage{xcolor}
\usepackage{verbatim}
\usepackage{graphicx}
\usepackage{esint}
\usepackage{epstopdf}
\usepackage{float}
\usepackage{bm}
\definecolor{darkblue}{rgb}{0,0,0.6}
\usepackage[colorlinks,linkcolor=darkblue,citecolor=darkblue,urlcolor=black]{hyperref}

\newcommand{\beq}{\begin{equation}} \newcommand{\eeq}{\end{equation}}

\begin{document}

\title{Modern computational studies of the glass transition}

\author{Ludovic Berthier}

\affiliation{Laboratoire Charles Coulomb (L2C), Universit\'e de Montpellier, CNRS, 34095 Montpellier, France}

\affiliation{Yusuf Hamied Department of Chemistry, University of Cambridge, Lensfield Road, Cambridge CB2 1EW, United Kingdom}

\author{David R. Reichman}

\affiliation{Department of Chemistry, Columbia University, 3000 Broadway, New York, New York 10027, USA}

\date{\today}

\begin{abstract}
The physics of the glass transition and amorphous materials continues to attract the attention of a wide research community after decades of effort. Supercooled liquids and glasses have been studied numerically since the advent of molecular dynamics and Monte Carlo simulations in the last century. Computer studies have greatly enhanced both experimental discoveries and theoretical developments and constitute an active and continually expanding research field. Our goal in this review is to provide a modern perspective on this area. We describe the need to go beyond canonical methods to attack a problem that is notoriously difficult in terms of time scales, length scales, and physical observables. We first summarise recent algorithmic developments to achieve enhanced sampling and faster equilibration using replica exchange methods, cluster and swap Monte Carlo algorithms, and other techniques. We then review some major recent advances afforded by these novel tools regarding the statistical mechanical description of the liquid-to-glass transition as well as the mechanical, vibrational and thermal properties of the glassy solid. We finally describe some important challenges for future research.        
\end{abstract}

\maketitle

\section{Introduction}

\label{sec:introduction}

When a liquid is quenched rapidly below its melting temperature it typically remains in a liquid state which, if cooled sufficiently, forms a glass~\cite{edigerangellnagel}. A glass is a solid that is as mechanically stable as many crystals yet is completely disordered, bearing none of the structural hallmarks of ordered periodic crystals. The cooling of a liquid to a glass spans approximately fifteen orders of magnitude in relaxation time. This dramatic slowing of dynamics is accompanied by a host of spectacular features, such as highly collective and non-exponential dynamics referred to as dynamical heterogeneity~\cite{dh}, a non-Arrhenius temperature dependence of the viscosity, and stark violations of the Stokes-Einstein relationship~\cite{ediger2000spatially}. The glass state itself differs in many ways from its crystalline counterpart, and a microscopic understanding of elementary excitations in the amorphous glass remains to this day, perhaps surprisingly, an open problem.

The liquid to glass transition shares deep analogies with similar phenomena in other fields of science, such as those exhibited by some magnetic systems (e.g., spin-glasses~\cite{binderyoung}), biological systems (e.g., protein folding and misfolding~\cite{bryngelsonwolynes}), and in computer science (e.g. satisfiability problems~\cite{mezardzecchina}). Understanding the process of glass formation and the factors that imbue disordered solids with their unique properties is thus justifiably considered one of the outstanding problems of condensed matter science. 

From a computational viewpoint, this task is plagued by several major obstacles, such as very sluggish dynamics, non-trivial finite size effects, ergodicity breaking, strong sample-to-sample fluctuations and self-induced heterogeneity typically shared by other complex systems~\cite{berthier2011theoretical}. Computer simulations provide a powerful means to probe the microscopic details of the dynamics, structure and thermodynamics of supercooled systems as the glass transition is approached~\cite{frenkelsmit,allentildesley}. However, such {\em in silico} experiments have historically been limited to small system sizes and the regime of very mild supercooling~\cite{kobandersen1}. The latter limitation is imposed by the fact that standard approaches require the local motion of particles to occur over temporal intervals that are much shorter than the time it takes the computer processor to carry them out, creating a bandwidth problem for slow dynamics. Thus, simulations of supercooled liquids have for a long time been confined to roughly the first five decades of dynamical slow down from the high temperature liquid, leaving the simulation and investigation of phenomena close to the experimental glass transition temperature $T_g$~\cite{edigerangellnagel}, as well as the realistic study of the glass itself, unapproachable.

In the last several years rapid progress has been made in the development and use of novel algorithms that allow researchers to circumvent this timescale bottleneck and enable the preparation of glassy states that are effectively cooled as slowly as those prepared in the laboratory~\cite{swap}. These techniques may even allow the simulation of glassy properties that are difficult to measure in real-world materials (e.g., growing static~\cite{bouchaudbiroli} and dynamical~\cite{franzparisi, berthier2007-1,berthier2007-2} length scales close to the glass transition), provide the ability to study novel phase transitions in glasses {\em in silico} for the first time (such as the brittle-to-ductile yielding transition~\cite{bonn, nicolas}), and afford the means to fill in the microscopic information absent from long-standing powerful but phenomenological theories (such as what actually tunnels in the two-level system model of Anderson, Halperin and Varma~\cite{ahv} and of Phillips~\cite{phillips}). Here, we outline the key methods and the breakthroughs that enabled them, as well as the myriad vistas on the nature of supercooled liquids and glasses that these techniques have opened.

\section{Computational methods}

\label{sec:computational}

\subsection{Basic tools and glass-forming models}

\label{sec:basic}

Computer simulations of glass-forming liquids employ molecular dynamics or Monte Carlo techniques both to generate equilibrated configurations under specified thermodynamic conditions and to calculate dynamical trajectories from these configurations~\cite{frenkelsmit,allentildesley}.  Molecular dynamics aims to mimic the true classical microscopic motion of particles, and is thus inherently local in terms of particle moves.  When used to simulate realistic dynamics, Monte Carlo is also constrained to be local in its exploration of configuration space~\cite{berthier2007monte}, otherwise the assignment of the time scale associated with particle moves is complicated and may not be possible~\cite{balnyets}.  On the other hand, for the generation of equilibrated configurations, Monte Carlo methods have the advantage that non-local and cluster moves may be employed, with the potential for greatly accelerated exploration of phase space and hence provide a more efficient generation of equilibrium configurations at high densities or low temperatures~\cite{luijten}.  Both techniques require the specification of the form of a interparticle potential energy function, as discussed below.  Molecular dynamics simulations then proceed via the calculation of the force between particles from this function, while Monte Carlo requires only the potential energy itself, and except in specialized approaches such as force-bias Monte Carlo,  does not generally require the calculation of forces.

Models of classical glass-forming systems can be crudely separated into three categories associated with the level of detail of the underlying description of the `particles' and the interactions between them.  The first category is that of realistic, off-lattice models of molecular glass-forming liquids.  Here, the goal is to model the microscopic details of the interactions between the most common glass-forming substances such as glycerol or silica.  The key component of such models is the form of the interaction potential or force field via which the atoms that form the molecules within the glass-forming liquid interact. Even here, distinctions may be made with respect to the degree of realism and detail associated with the model. For example, the van Beest, Kramer, van Santen BKS model of silica~\cite{bks} is defined by a potential energy surface which has been parametrized from a combination of {\em ab initio} and experimental data, and consists of Coulomb, Born-Mayer repulsive, and dispersive interactions between the Si and O atoms.  On the other hand, the Lewis and Wahnstr\"om model of the organic molecular liquid ortho-terphenyl replaces the entirety of each of the three benzene rings of the molecular unit with a single site which interacts with other sites with a simple Lennard-Jones potential~\cite{lewiswahnstrom}. The latter case clearly involves some degree of atomic coarse-graining while retaining the non-spherical, albeit rigid, structure of the ortho-terphenyl molecular unit.  The simulation of the glass-forming behavior of all members of this category is much more intensive than for either of the simpler categories described below.  This is because the treatment of periodic systems composed of units with non-spherical shapes and rotational degrees of freedom, long-ranged interactions, and other features poses complications for standard molecular dynamics and Monte Carlo simulations that render simulations more time consuming than for simpler models where these features are absent~\cite{frenkelsmit}.  

The models in the second category of {\em in silico} glass-formers describe simple systems of spherical particles interacting with short-ranged interactions.  These systems may have a potential energy function that is purely repulsive, such as the hard-sphere potential, or somewhat more complex interactions such as those described by the Lennard-Jones potential~\cite{weber,kobandersen1,kobandersen2}. Generally some degree of polydispersity and a tuning of the interactions are needed to prevent facile crystallization.  While the potential energy functions for the models in this class may be employed to describe experimentally relevant glass-formers, such as colloidal particles or some specific metallic glasses~\cite{dh}, the standard philosophy for their use is the fact that they demonstrate a nearly full range of non-trivial behaviors exhibited by more complex molecular glass-formers while being much more efficient to simulate numerically. In this sense, systems in this category form a middle ground between the complexity of the molecular systems discussed above, and the fully coarse-grained models described below.

Finally, we come to models in the third category, which are fully coarse-grained lattice models.  These models place particles on a lattice, with simple thermodynamic constraints imposed, such as restrictions on the number of neighbors allowed for a given particle type.  There are no forces acting on particles, and dynamical evolution occurs via local Monte Carlo moves~\cite{birolimezard}.  These models, sometimes referred to as `lattice glass models', generally suffer from a tendency to easily crystallize, although recently progress has been made in the design of simple lattice models for which crystallization is strongly frustrated~\cite{hukushima}.  The main utility of lattice glass models resides in the fact that they are simple enough to directly apply powerful mathematical techniques such as the replica method~\cite{mpv} for the calculation of thermodynamic properties to their simple lattice energy functions, and thus they form a relevant bridge between theories of the glass transition and more complex, off-lattice models.

In this review we will largely focus on models in the second category.  These models provide an excellent compromise between the complexity of molecular systems and lack of realism of lattice models.  They are realistic enough to be viewed as reasonable proxies for the simplest laboratory glass-formers such as metallic glasses, and thus we may view their study on the computer as {\em in silico} experiments on such systems.  In particular, given the fact that the experimental behavior that computer simulations aim to capture, such as violation of the Stokes-Einstein relationship, dramatically growing time scales, and the appearance of non-exponential and non-Arrhenius relaxation, are shared by a diverse set of molecular laboratory glass-formers and computer simulations of supercooled spherical particles with short-ranged interactions, it is reasonable to believe that little, if anything of physical importance is being excluded by focusing on their study.  Of course properties  specific to non-spherical degrees of freedom, such as the rotational version of the Stokes-Einstein relation or realistic modeling of dielectric relaxation cannot be described by these models~\cite{dh}, and we will not discuss such properties here.

\subsection{Equilibration tools in complex systems}

\label{sec:equilibration}

A large portion of this review will focus on the simulation techniques used to equilibrate glassy samples on the computer.  It is here that some of the biggest challenges exist for the computer simulation of glassy materials.  Using local moves to produce equilibrated particle configurations is plagued by the time scale issues associated with dynamical slowing down. The increases in processor speed, even those afforded by newer architectures such as GPUs~\cite{gpu}, are insufficient to enable the equilibration of samples anywhere close to the thermodynamic location of glass transition itself. Thus, tailored algorithms making use of non-local Monte Carlo and related techniques must be employed.  It is also here that the greatest recent progress has been made.  This progress and the myriad applications it has enabled, forms the core of this review. Because relaxation timescales controlling particle motion change by many orders of magnitude as temperature decreases, an algorithmic speedup of one or two orders of magnitude as often encountered below, although useful and welcome, does not offer radical changes to the probed physics.

It is useful to briefly sketch the history of the development of methods for the equilibration of the Ising model~\cite{ising} and its cousins, such as binary alloys and spin glass models~\cite{mpv}, as a means both to illustrate the introduction of approaches to deal with slow equilibration issues, e.g. the slowing down near a critical point, as well as distinctions between the requirements for an efficient algorithm in these simpler models and those associated with the off-lattice problem of glassy liquids. 

Standard single spin flip Monte Carlo approaches were first employed to simulate the thermodynamics of the Ising model soon after the introduction of the Monte Carlo algorithm in 1953~\cite{teller}.  These approaches encountered severe problems as the critical point is approached and equilibration times scale as a power of the growing length scale associated with emergent ferromagnetic order~\cite{suzuki}.  Standard local Monte Carlo was also attempted around the same time to study order-disorder phenomena in binary alloy models.  In 1959, Fosdick~\cite{fosdick}, as well Salsburg {\em et al.}~\cite{salsburg} introduced non-local swap moves into the Monte Carlo framework for such systems.  Interestingly, this approach took more than a decade to find its way into the study of off-lattice models of liquids, and several decades more to find applications in the study of glasses.  As we will discuss in this review, a modern version of this idea, as embodied in the swap Monte Carlo (SMC) approach, has recently revolutionized the study of models of glass-formers belonging to the second class described above.
In the next subsection we will discuss cluster and swap-based Monte Carlo techniques, but first we discuss an alternate route to accelerated equilibration.

\begin{figure*}
    \includegraphics[width=16.5cm]{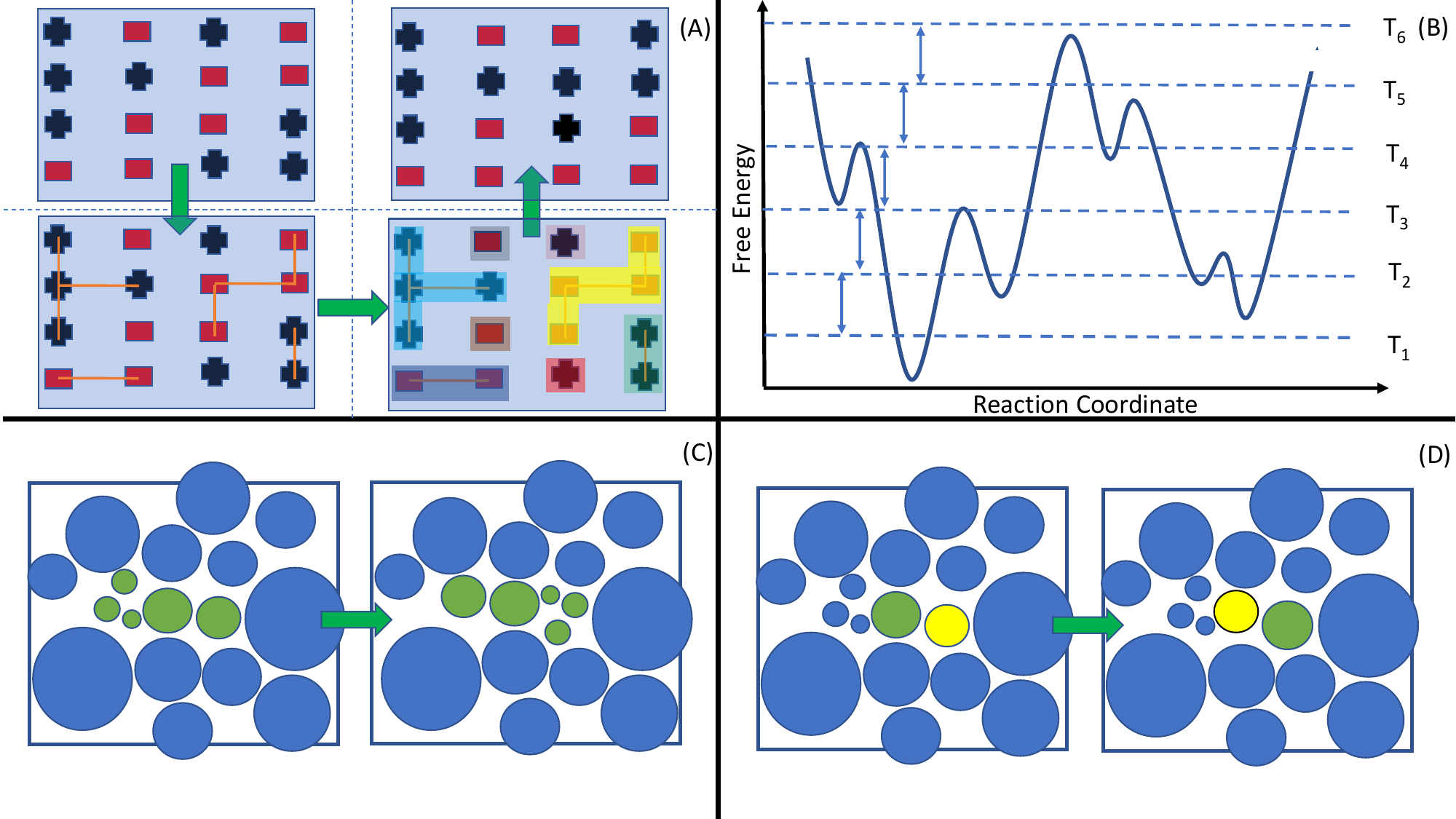}
    \caption{Sketch of various equilibration methods. (A) Swendsen-Wang cluster Monte Carlo for lattice models, illustrated with the example of a random-coupling Ising model in 2D. First, bonds are specified between like ``up" or ``down" spins depending on their coupling strength. This defines clusters of spins that may be flipped with a Monte Carlo probability. Concept adapted from Ref.~\cite{luijten2}. (B) Cartoon illustrating parallel tempering Monte Carlo. Monte Carlo calculations are run at different temperatures (here 6) and the temperatures or configurations exchanged, allowing for a more rapid sampling of phase space. (C) Cartoon of a cluster move in the Monte Carlo algorithm of Dress and Krauth. Adapted from Ref.~\cite{santen} (D) Cartoon of a swap Monte Carlo move.}
    \label{fig:sketch}
\end{figure*}

In 1986 Swendsen and Wang introduced the replica Monte Carlo method to study the thermodynamics of Ising spin-glasses~\cite{swendsen1}. Here, replicas of the system at different temperatures are simulated in equilibrium, with a partial exchange of configurational information allowed between replicas.  A related but somewhat distinct and more general approach was put forward several years later by Geyer~\cite{geyer} and by Marinari and Parisi~\cite{mandp}, who christened the approach simulated tempering and used it to investigate the random-field Ising model. In simulated tempering, a set of independent systems are simulated in equilibrium, with a Metropolis-like Monte Carlo exchange of temperatures between equilibrium configurations which maintains equilibrium, see Fig.~\ref{fig:sketch}. This class of Monte Carlo approach, which is frequently called parallel tempering or replica-exchange Monte Carlo~\cite{hukushima1996exchange}, has been of tremendous utility in the simulation of complex systems with rough energy landscapes in fields ranging from materials science to biology~\cite{deem}.  For the study of bulk glasses, its performance has been shown to be modest, although we will discuss important applications where it is currently the most efficient method available.

A variety of other methods have been invented and carefully investigated for the equilibration and simulation of lattice models.  Some of these methods appear quite powerful.  The use of these approaches for supercooled liquids, however, has been sporadic and the results obtained for glassy systems somewhat anecdotal~\cite{faller1,faller2,trebst,bogdan}.  While these techniques merit further investigation, they will not be discussed further here.  

Enhanced sampling methods such as replica-exchange and related techniques have existed in the literature for several decades and have been applied very successfully to the sampling of rough energy landscape problems in fields that bare some relationship to the glass transition problem, such as spin-glasses, the conformational equilibrium of polymers and biomolecules, and the screening of low-energy crystal structures~\cite{deem}.  The situation is somewhat different for the study of glassy systems. To illustrate this, we first focus on the application of a molecular dynamics version of replica-exchange, called replica-exchange molecular dynamics (RXMD), to equilibrate off-lattice supercooled liquid configurations  by Yamamoto and Kob~\cite{yamamoto2000replica}.

Yamamoto and Kob explored RXMD in the binary Lennard-Jones Kob-Andersen model, perhaps the most ubiquitous of the second class of models as categorized in Sec.~\ref{sec:basic}, and followed the protocol of Sugita and Okamoto in the implementation of RXMD~\cite{sugitaokamoto}.  In particular, they employed $M$ non-interacting replicas of systems of $N$ particles each, where the Hamiltonian of each replica has its potential energy function scaled by a constant parameter acting as a means of controlling the temperature in the configurational average of the replicas. A constrained molecular dynamics simulation was then performed on the entire system comprised of all replicas.  Lastly, the exchange of the scaling parameters (effectively an exchange of temperatures) between distinct replicas was considered on a specified time interval, with acceptance rate governed by the standard canonical Metropolis criterion.  This scheme is guaranteed to lead to canonical equilibrium at the set of temperatures of the replicas, and reverts to standard molecular dynamics when no exchanges are attempted.

The efficient use of enhanced sampling techniques requires fine-tuning of the algorithmic parameters. In their simulations, Yamamoto and Kob used 16 replicas of $N=10^3$ in a range in temperature from the onset of glassy dynamics to the mode-coupling crossover temperature $T_{c}$, covering approximately 4 decades of slowing down. The mode-coupling temperature marks the border of a regime where glassy behavior becomes quite difficult to directly simulate in equilibrium via local molecular dynamics or Monte Carlo, even for current-day computers.  As expected, the efficiency of the RXMD algorithm was found to strongly depend on the exchange time scale, which was optimized and found to be on the order $10^3$ time steps.  Physically, this corresponds to the approximate time scale of oscillations of a particle trapped by its neighbors. Via monitoring the effective diffusion constant of the particles, Yamamoto and Kob concluded that RXMD was up to 100 times more efficient at exploring phase space in the regimes studied, although equilibration below $T_c$ was not attempted.

De Michele and Sciortino revisited this approach in a simpler model of a one-component Lennard-Jones liquid with a term added to inhibit crystallization~\cite{sciortino}. They noted that the effective diffusion constant within replica exchange-based approaches is controlled by the diffusion in the highest temperature replica, and carries little to no information on the equilbration rate. Instead, De Michele and Sciortino focus on the time scale to find the lowest local minimum in the potential energy landscape~\cite{stillingeris}. Within the purview of this more stringent criterion, De Michele and Sciortino find that using RXMD is not more efficient than standard molecular dynamics for equilibrating supercooled particle configurations.

While it should be remarked that other authors have also employed variants of replica exchange or parallel tempering as a means to accelerate the sampling of equilibrium glassy configurations, these studies often combine the approach with other sampling techniques, making it difficult to isolate the role played by the replica technique itself~\cite{faller2}.  It is important to note that while RXMD, parallel tempering, and related techniques do not appear to afford dramatic increases in equilibration efficiency for bulk supercooled liquids, they do appear to provide the most efficient means of equilibrating supercooled liquids in confined geometries and in situations where a fraction of the particles are artificially frozen in place.  Such situations are of vital importance for the extraction of growing static length scales as the glass transition is approached.  These applications, and the power of replica-based approaches for studying them, will be discussed in Sec.~\ref{sec:statmech}.

So far, we discussed computational tools that are widely used in different physical situations to study supercooled liquids. We close this section with two approaches that were proposed and developed more specifically for glassy systems.

In 2007, Ediger and coworkers discovered that amorphous films prepared using physical vapor deposition in well-chosen conditions had properties nearly equivalent to bulk glasses prepared at exceedingly small rates~\cite{swallen2007organic}. After 15 years of detailed studies~\cite{ediger2017perspective}, it is understood that physical vapor deposition represents an experimental approach to accelerate the equilibration of supercooled liquids, with a speedup that can reach many orders of magnitude. The physical origin of this observation is also understood: molecules arriving at the surface of the film have a much larger mobility than the ones already buried in the bulk. This enhanced surface mobility allows them to easily relax at temperatures where the bulk is arrested. Motivated by this discovery, algorithms mimicking the deposition process were developed~\cite{singh2013ultrastable,lyubimov2013model}. However, on the timescales accessible to computer simulations, surface and bulk dynamics differ by at most 1-2 orders of magnitude, and the speedup afforded by this method is much smaller than in experiments~\cite{berthier2017origin}. In addition, simulating the growth process itself is not straightforward and equilibration is not guaranteed. Therefore, simulating vapor deposition is useful to help interpret experimental studies~\cite{dalal2015tunable}, but is not a promising generic tool to speedup equilibration.    

The second method is more theoretically guided. The initial goal was to understand the nature of large deviations in the dynamic behaviour of glass-formers~\cite{merolle2005space} as a way to describe more formally the nature of dynamic heterogeneity~\cite{chandler2010dynamics}. Technically, the idea is to introduce a non-equilibrium sampling technique which biases the system towards dynamic trajectories exhibiting statistically rare properties, such as for instance lower than average mobility~\cite{garrahan2009first}. An outcome is the production of particle configurations that have physical properties that are different from the bulk, and appear to lie deeper in the potential energy landscape than equilibrium systems at the same temperature~\cite{jack2011preparation,keys2015using,turci2017nonequilibrium}. There is ample evidence that these configurations represent very stable glassy configurations, but this has not been quantified. In addition, these tools do not scale well with system size, and are currently limited to relatively small systems. This is a promising technique whose utility and performances remain to be more quantitatively established.

\subsection{Advanced Monte Carlo techniques: Cluster Monte Carlo and swap Monte Carlo}

\label{sec:advanced}

In 1987, Swendsen and Wang introduced an extremely powerful approach for simulating Ising-type systems close to criticality~\cite{swendsen2}. The essence of this method is the use of non-local but detailed-balance-preserving Monte Carlo cluster moves, see Fig.~\ref{fig:sketch}.  Swendsen and Wang illustrate this approach in both the two-dimensional Ising model as well as Potts models, where the critical slowing down associated with the second-order critical point is greatly mitigated by the algorithm's violation of dynamical universality, rendering a much less severe space-time scaling exponent, which translates into a much more efficient algorithm when compared to standard single-spin Monte Carlo moves.  Interestingly, as we will discuss below, the efficiency of the SMC algorithm manifests in a related manner in the simulation of glassy liquids.  An analog of the Swendsen and Wang cluster approach for simple off-lattice liquids and glasses was devised by Krauth and co-workers~\cite{dress}.  This approach will be discussed in more detail below.  The current limitation of cluster approaches for off-lattice models lies in the difficulty of efficiently determining and moving clusters.  This is a much more challenging task than in the Ising model, where, for example, the physics revealed by the knowledge of Fisher clusters~\cite{fclusters} and the Fortuin-Kasteleyn~\cite{FK} representation, greatly simplifies the construction of the algorithm.

The pioneering work by Swendsen and Wang on spin models provides an impetus for the search for efficient cluster Monte Carlo approaches to accelerate the equilibration of particle-based glassy systems. In 1995, Dress and Krauth~\cite{dress} devised a cluster approach in which a copy of a particle configuration is rotated with respect to the original configuration as a means to identify clusters in the joint system via an overlap criterion, see~Fig.\ref{fig:sketch}. Each cluster can then be flipped around a pivot via a Metropolis procedure in which spheres belonging to one cluster are moved from the rotated copy back to the original configuration while those in the original configuration are moved to rotated copy. Such moves, in conjunction with simple single-particle Monte Carlo moves, satisfy detailed balance and can potentially accelerate the exploration of configuration space.

Five years after the publication of the cluster Monte Carlo algorithm of Dress and Krauth, Santen and Krauth used this approach to study glass formation in a polydisperse hard disk system in two dimensions~\cite{santen}.  Here, cluster Monte Carlo appears to be clearly more efficient than local Monte Carlo when judged by the reasonable criteria that the cluster approach enables the equilibration of configurations well past the mode-coupling density of the system, where a power-law fit of the diffusion constant would predict vanishing particle diffusivity.  These authors emphasize that even at the highest densities studied, their system displays no evidence of a change in thermodynamic quantities such as the compressibility which could be interpreted as an equilibrium signature of a glass transition. However, the putative location of a thermodynamic transition for this system is unknown, and so these results merely prove that the mode-coupling transition does not correspond to a thermodynamic singularity. In any case, even in theories that purport the existence of a thermodynamic glass transition such as the random first-order transition (RFOT) theory, such a transition is not expected to exist in two dimensions~\cite{RFOT}.  As discussed further below, more recent simulations using SMC have provided stronger evidence that no signatures of a thermodynamic transition exists in two dimensions but do exist in three dimensions.  

The cluster Monte Carlo algorithm of Krauth has been used far less extensively than SMC, which we describe next.  For simple systems of the second class of models, such as the hard disks studied by Santen and Krauth, it can be demonstrated that SMC is a more efficient algorithm~\cite{brumer}. Some of the difference in the efficiency of these two approaches likely lies in the relative simplicity of the swap approach, which allows for a significantly greater ease of optimization.  In particular, the algorithm defined by Dress and Krauth merely defines one possible approach for isolating and exchanging clusters. Indeed, it should be noted there are more general versions of the algorithm of Dress and Krauth, which have not as of yet been applied to glassy systems~\cite{liuluijten}.  It is likely that related algorithms could be devised which would exceed the capabilities of the Dress and Krauth approach, and that combining the cluster algorithm with other novel sampling techniques could greatly enhance its efficiency.  One example of work in this direction is the use of the cluster Monte Carlo approach with the rejection-free event-chain Monte Carlo approach, which indeed demonstrably enhances the sampling ability of cluster Monte Carlo~\cite{bernard2009event,krauth2021event}. Currently, for models of complex glass-forming liquids, the most powerful means of achieving deeper supercooling is via direct molecular dynamics simulation, making use of high-throughput methods~\cite{simmons} or using advanced special-purpose hardware architecture such as Anton~\cite{shaw}. Thus, in our opinion the future search for such potential modifications is a worthy goal due to the somewhat circumscribed set of systems where SMC is extremely useful.  Indeed, for extremely simple systems such as the monodisperse glass former studied by De Michele and Sciortino, SMC trivially affords no advantage, while for complex molecular liquids of the first class of models we have described, it appears to be quite difficult to apply.  

Above, we have casually compared the cluster Monte Carlo approach to the SMC method without formally defining the latter, mostly because the approach is nearly self-explanatory.  Here, we touch upon the history of the approach, some details associated with its implementation, and the steps leading to its great success as a means of generating deeply supercooled liquid configurations.

In the SMC algorithm, standard local Metropolis Monte Carlo is augmented with the potentially long-ranged swapping of pairs of particles, see~Fig.\ref{fig:sketch}. Optimization must be carried out with respect to the frequency of swapping trial moves and the range of particle sizes and types for which swap moves are attempted, but since all moves occur via the Metropolis criterion, the method is extremely simple and requires very little tuning.  In systems where the approach is efficient, swap moves may be rarely accepted, but when they are they can provide an enormous boon for equilibration. The degree to which the approach is useful will depend very sensitively on the acceptance rate of swap moves compared to local moves, which will vary greatly from system to system.

A variant of this type of Monte Carlo was used to study binary crystalline alloys a mere six years after the invention of the basic Monte Carlo algorithm itself. The first off-lattice use of SMC was carried out by Tsai, Abraham and Pound in 1978, who used the approach to investigate the structure and thermodynamics of relatively small binary Lennard-Jones clusters~\cite{tap}.  In 1989, Gazzillo and Pastore used SMC to investigate the equation of state of non-additive hard-sphere mixtures~\cite{gp}.  It thus took a full 30 years from the time of invention for the approach to be employed to study bulk liquids.

In a pioneering work, Grigera and Parisi employed SMC for the study of the glassy behavior of a 50:50 mixture of soft-spheres~\cite{grigeraparisi}.  Although no detailed statements were made in this work concerning the acceleration of equilibration over standard molecular dynamics or Monte Carlo, these authors demonstrated that at high densities or low temperatures (these parameters are equivalent in soft-sphere systems) the inclusion of swap moves renders the algorithm much more efficient at reaching low energy configurations on the energy landscape compared to both standard Monte Carlo and parallel tempering. Grigera and Parisi applied SMC to a system of 34 spheres, showing that a rather stringent metric of equilibration, namely the agreement between the specific heat as calculated from energy fluctuations and the temperature derivative of the average energy, holds down to temperatures well below the expected location of the laboratory glass transition.  They interpreted the broad maximum of the specific heat in this small mixture with the location of an entropy crisis. Via extrapolation, they illustrate a similar but sharper behavior in an 800 particle system, with the location of the peak in the specific heat roughly agreeing with the predicted Kauzmann temperature as found from RFOT theory~\cite{RFOT,mezardparisi}.

\begin{figure}
    \includegraphics[width=0.9\columnwidth]{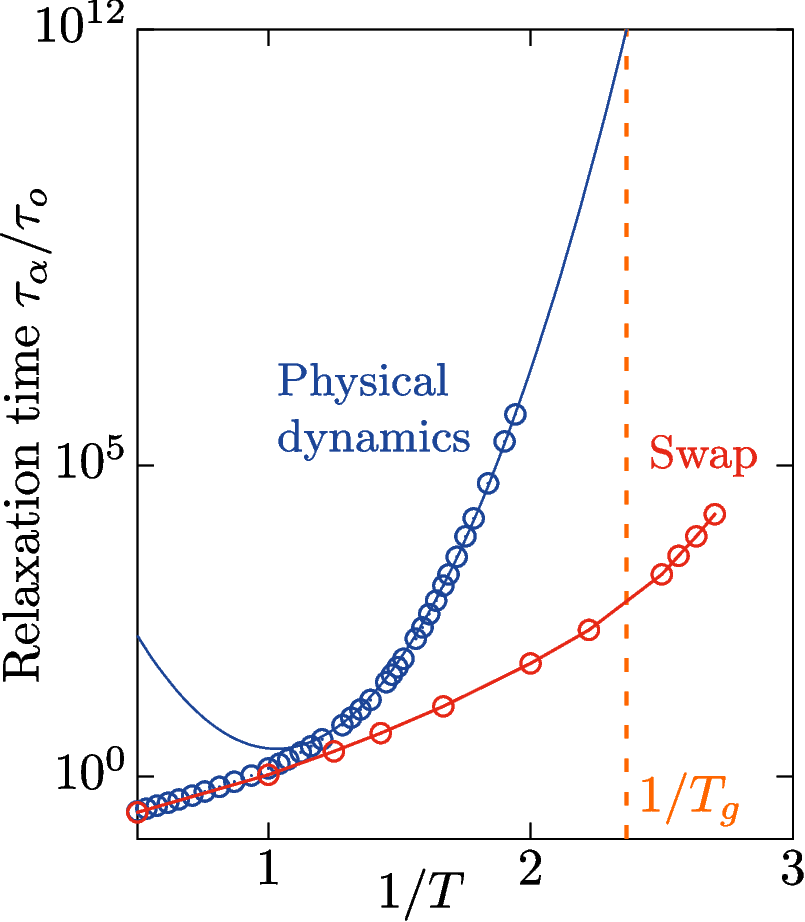}
    \caption{Illustration of the giant equilibration speed up offered by swap Monte Carlo (red) versus physical dynamics (blue) for a three-dimensional system of polydisperse soft spheres. The equilibration is about $10^{10}$ at the experimental glass transition $T_g$ and equilibration can be achieved below $T_g$ to produce ultrastable computer glasses. Adapted from Ref.~\cite{swap}.}
    \label{fig:dynamics}
\end{figure}

The status of the SMC approach to equilibrate supercooled particle configurations after the work of Grigera and Parisi remained unclear for several years.  Here, two important studies stand out.  Fernandez {\em et al} attempted to quantify the degree of acceleration of equilibration with SMC in the same system employed by Grigera and Parisi. These authors found that SMC accelerated equilibration by approximately two orders of magnitude independently of temperature, a speed up which is helpful but far from sufficient to study realistically-annealed samples~\cite{fernandez}.  Brumer and Reichman investigated SMC in several systems, including the polydisperse two dimensional hard disk system invented by Santen and Krauth~\cite{brumer}. They found that for this system, SMC was more efficient than the pivot-based cluster Monte Carlo approach.  However, less encouragingly, Brumer and Reichman concluded that SMC suffered from a proclivity to crystallize all other systems under investigation, including three-dimensional thermal analogs of the hard disk system of Santen and Krauth.

Perhaps because of these negative results, progress in the use of SMC stalled until the important work of Ninarello, Berthier and Coslovich in 2017~\cite{berthier2016efficient,swap}. These authors showed that if the interaction potential and the polydispersity of the sample are carefully tuned, amazingly efficient equilibration in large configurations is possible without signs of crystallization. The challenge was to develop models for which swap moves are easily accepted, which can be achieved using either discrete mixtures where swap between different particle families can be performed, or continuously polydisperse systems. Because such polydisperse models can however easily crystallise, particle interactions need to be carefully adjusted to prevent fractionation or phase separation, using for instance non-additive pair interactions. When these conditions are met, the efficiency of the approach is so high that one can easily reach and exceed the degree of annealing found in standard laboratory protocols, as illustrated in Fig.~\ref{fig:dynamics}. In the intervening years a host of model systems amenable to remarkably efficient equilibration via SMC have been devised~\cite{swap,lammpsswap}. Recent work has also demonstrated that models of metallic glasses such as the venerable Kob-Andersen model~\cite{parmar2020ultrastable}, which mimics the NiP metallic glass-forming system, can be well approximated by a potential for which SMC can be efficiently carried out. This work has enabled a large number of previously impossible investigations into the properties of supercooled liquids and glasses in the second category of models as defined above. The remainder of this review in Sec.~\ref{sec:recent} will discuss this progress.

The performance demonstrated in Fig.~\ref{fig:dynamics}, which represents a speedup of a factor larger than $10^{10}$ at the experimental glass transition $T_g$, appears surprising in the context of advanced Monte Carlo techniques. In the cases discussed above, a computational bottleneck rooted in the physics of the problem was tackled using a technique originating from physical intuition. In the case of supercooled liquids, the intuition that a rugged energy landscape controls the physics would favour methods like parallel tempering which do not work well. The real space view associating slow dynamics to some form of spatial correlation between particles~\cite{dh,bouchaudbiroli} would instead suggest the need for collective cluster moves. In this second view, the success of swap Monte Carlo which introduces very basic two-particle moves is surprising. These considerations have led to several studies confronting the speedup offered by SMC to the physics of supercooled liquids~\cite{wyart2017does,ikeda2017mean,szamel2018theory,berthier2019can}. Physically, the key resides in the interplay between the translational degrees of freedom (particle positions) and the diameter dynamics introduced by the swap exchanges~\cite{swap}. The idea of augmenting the number of degrees of freedom has led to novel algorithms that have proven useful in the context of the jamming transitions~\cite{brito2018theory,hagh2022transient}.

\section{Recent advances in understanding the glass problem}

\label{sec:recent}

\subsection{Statistical mechanics analysis of glass transition}

\label{sec:statmech}

The equilibration speed up afforded by SMC permits the production of a large number of independent equilibrium configurations of the glass-former under study over a temperature regime that encompasses the experimental glass transition temperature. This naturally provides a means to perform ensemble-averaged measurements of any equal-time correlation function and, by integration, any thermodynamic quantity of physical interest. 

Since the landmark work of Kauzmann~\cite{kauzmann1948nature}, the configurational entropy $S_{\rm conf}(T)$ of supercooled liquids has played a special role in glass studies~\cite{berthier2019configurational}. Gathering available experimental data, Kauzmann provided estimates for the temperature dependence of $S_{\rm conf}$ and noticed a steep decrease as temperature decreases towards $T_g$. Extrapolating this evolution to temperatures below $T_g$ where no experimental data is available, Kauzmann noted the possibility that a critical temperature, now known as the Kauzmann temperature $T_K$, could mark an entropy crisis with $S_{\rm conf}(T \to T_K)=0$. Theoretical developments have since greatly clarified the conceptual, mathematical and physical contents of the configurational entropy~\cite{berthier2019configurational}. In the mean-field theory of the glass transition describing the physics of supercooled liquids in the limit of a large number of spatial dimensions, $d \to \infty$, a Kauzmann transition accompanied by a vanishing configurational entropy rigorously exists~\cite{parisi2020theory}. In this framework, which serves as a basis to the RFOT theory~\cite{wolynes2007review}, $S_{\rm conf}$ quantifies the complexity of a rugged free energy landscape with a clear mathematical definition that does not involve any reference to a crystalline state.   

It is not yet known whether a Kauzmman transition can exist in finite dimensions, $d<\infty$, but some key mean-field concepts are known to be dramatically affected by finite dimensional effects~\cite{biroli2014random}. In particular, it is impossible to simply and rigorously define, let alone enumerate, long-lived free-energy minima and the mean-field definition of $S_{\rm conf}$ must be carefully reconsidered~\cite{RFOT}. About 20 years ago, a series of numerical works following older ideas by Goldstein~\cite{goldstein1969viscous} and Stillinger and Weber~\cite{stillingeris} introduced a definition of the configurational entropy based on potential energy (rather than free energy) minima~\cite{sciortino1999inherent,sastry2001relationship}. Although this was known to be an approximation, it permitted the development of explicit computational methods to obtain an estimate of $S_{\rm conf}(T)$ which has been employed across a wide range of models. A strong limitation to these early efforts is the narrow temperature range covered by these measurements, which is mostly above the mode-coupling crossover temperature and corresponds, within the RFOT theory, to a regime where $S_{\rm conf}$ cannot even be defined.

\begin{figure}
    \includegraphics[width=\columnwidth]{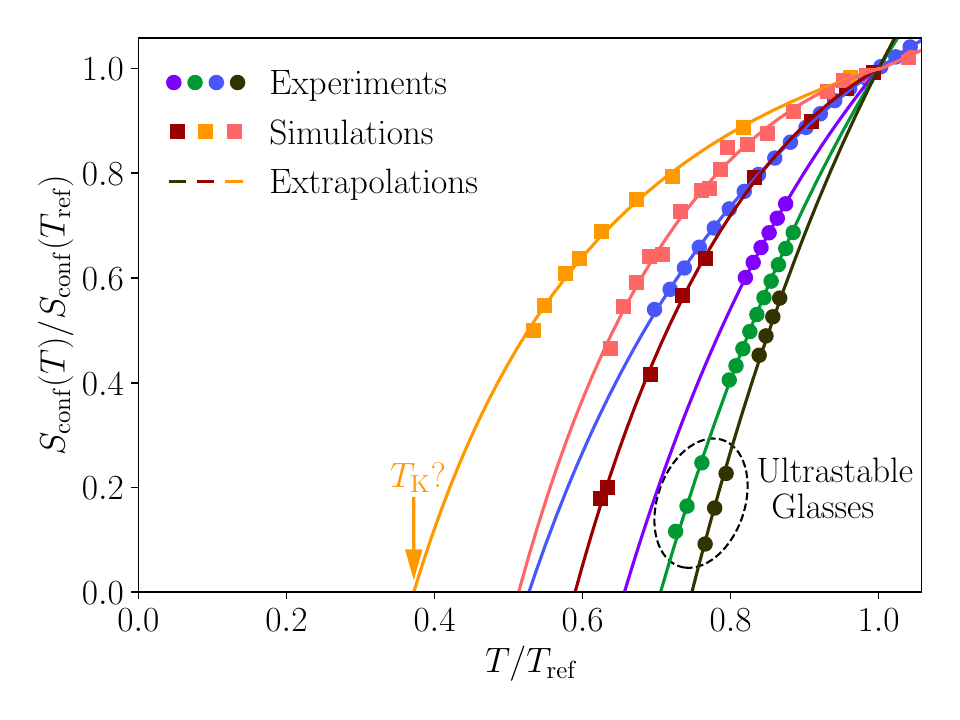}
    \caption{The evolution of experimental (including ultrastable thin films) and numerical estimates (obtained using the swap Monte Carlo approach) of the configurational entropy offers a consistent view of the thermodynamics of three-dimensional glasses. This data compilation, adapted from Ref.~\cite{berthier2019configurational}, revisits the famous plot first constructed by Kauzmann~\cite{kauzmann1948nature} for the excess entropy of molecular supercoooled liquids.}
    \label{fig:kauzmann}
\end{figure}

The situation changed after 2017 when the SMC algorithm suddenly opened a path to analyse the thermodynamic properties of supercooled liquids down to $T_g$ and even below. At the methodological level, new methods were developed to provide computational estimates of $S_{\rm conf}$ that are conceptually much closer to the rigorous theoretical definition of this quantity~\cite{berthier2014novel,ozawa2018configurational} while in practice these measurements could now be performed in the temperature regime where theory predicts its validity and experimental estimates exist. This new generation of measurements were performed across a range of simple models of the second category described in Sec.~\ref{sec:basic} in both two and three dimensions~\cite{berthier2019configurational}. The first result is that the steep decrease of $S_{\rm conf}(T)$ reported by Kauzmann is recovered in all models~\cite{berthier2017configurational}, see Fig.~\ref{fig:kauzmann}. This is not trivial given the diversity of molecules analysed in Kauzmann's work and the different nature of the quantity he reported. This confirms in particular that no reference to the crystalline state of the material is needed to estimate $S_{\rm conf}$. More quantitatively, extrapolating the numerical data to temperatures where even SMC is unable to provide equilibrated configurations suggests that a finite Kauzmann temperature $T_K>0$ can exist in three-dimensional models~\cite{berthier2017configurational}, whereas a different behaviour is found in two dimensions where extrapolations seem to suggest that $T_K=0$~\cite{berthier2019zero}. These findings are in line with RFOT theory, which implies that the Kauzmann transition should be destroyed by finite dimensional fluctuations in $d \lesssim 2$. 

These measurements, together with theoretical developments, give additional insight into the nature of the putative Kauzmann transition. Within the mean-field description, the entropy crisis at $T_K$ corresponds to a first-order change between a metastable glass phase above $T_K$ to an `ideal' glass phase below $T_K$ with vanishing configurational entropy. For $T > T_K$ which is the regime explored in equilibrium conditions, the glass phase is metastable with respect to the liquid, and the configurational entropy can be interpreted as the free energy difference between the two phases~\cite{franz1997phase}. This has two interesting consequences for simulations. First, it provides a computational path~\cite{berthier2014novel} to estimate the configurational entropy using free energy calculations of the type developed to analyse conventional phase transitions~\cite{frenkelsmit}. This typically requires measuring large deviations in the fluctuations of the order parameter, and methods such as umbrella sampling are well suited for such tasks and they can be readily adapted to the case of supercooled liquids~\cite{berthier2013overlap}. 

A second consequence of the glass metastability above $T_K$ is the possibility to induce a discontinuous phase transition towards the glass phase by application of a thermodynamic field that favors the glass, usually denoted as $\epsilon$~\cite{franz1997phase}. This amounts to adding a new dimension to the equilibrium phase diagram of supercooled liquids. In this extended phase space $(T,\epsilon)$, the Kauzmann transition signals the liquid-glass phase change when $\epsilon=0$ but a discontinuous first-order transition line $\epsilon(T)$ emerges from the Kauzmann point for temperatures above $T_K$. This line ends at a second-order critical point $(T_c, \epsilon_c)$ which RFOT theory predicts should lie in the same universality class as the random field Ising model~\cite{biroli2014random}. From an experimental viewpoint, these considerations may appear as formal theoretical developments. However, they are directly amenable to quantitative numerical tests, a program that was started about twenty years ago~\cite{franz1998effective,cardenas1999constrained}. However, these initial studies turned into quantitative tests only after SMC was developed. A complete exploration of the $(T,\epsilon)$ phase diagram, together with finite-size scaling analysis of the corresponding phase transitions are now available~\cite{berthier2015evidence,guiselin2020random,guiselin2022statistical}. These studies fully confirm the existence of a first-order transition line in the regime $T>T_K$ for three-dimensional glass-formers, and scaling analysis confirms the universality class of the critical end-point~\cite{guiselin2020random}. In two-dimensional models, no critical end-point is found, which agrees with studies of the random field Ising model itself, and scaling analysis again demonstrates good agreement with a zero-temperature Kauzmann transition~\cite{guiselin2022statistical}. Overall, these thermodynamic results are strong hints that the random first order transition theory of the glass transition provides an accurate description of the static properties and thermodynamic fluctuations in supercooled liquids. However, similar fluctuations and behaviour can be generically expected in systems displaying growing static order, such as multi-spin plaquette models~\cite{garrahan2002glassiness}, which exhibit constrained phase transitions as well~\cite{jack2016phase}, but are of course devoid of any finite temperature Kauzmann transition. 

Because the Kauzmann transition is discontinuous, no critical fluctuations of an order parameter are expected to grow as $T_K$ is approached from above, even in the mean-field limit. Therefore, the search for growing length scales as a sign of emerging order is more complicated for glasses than it is for simpler types of phase transformations. Generally speaking, the length scale which is growing in the vicinity of a first-order transition is a nucleation length scale, usually defined as the critical size that a nucleus of the stable phase must have in order to destabilise the metastable one~\cite{frenkelsmit}. In the context of the glass transition, this analogy has been employed to rigorously define~\cite{montanari2006rigorous} the corresponding length scale, now called the point-to-set length scale. A practical algorithmic procedure was also proposed~\cite{bouchaudbiroli} to measure the point-to-set length scale. Since the free energy difference between glass and liquid phases above $T_K$ is directly related to $S_{\rm conf}$, the point-to-set length scale is expected to be inversely proportional to this free energy driving force given by $S_{\rm conf}$, which would possibly diverge at $T_K$ where the entropy vanishes. 

In practice, this algorithmic construction works as follows. The positions of all particles outside a spherical cavity are frozen in an equilibrium configuration to impose the glass metastable phase outside the cavity. Particles inside the cavity are let free to evolve, and will eventually relax (thus returning to the liquid phase) when the cavity size becomes larger than the critical nucleation radius. By monitoring the typical state of the interior of the cavity as a function of the cavity size, a characteristic point-to-set length scale can be measured numerically~\cite{cavagna2007mosaic,biroli2008thermodynamic,hocky2012growing}. In practice, particles inside the cavity are so strongly constrained by the frozen boundaries that it is difficult to probe their thermodynamic properties even with SMC. An additional equilibration effort involving parallel tempering is needed to properly measure the point-to-set length scale~\cite{berthier2016efficient}. The agreement with the temperature evolution of $S_{\rm conf}$ was confirmed~\cite{berthier2017configurational,ozawa2019does}, directly demonstrating how a decreasing entropy, a growing point-to-set length scale and a decreasing free energy difference between glass and liquid all reveal the proximity to a Kauzmann transition and can be detected in simulations of bulk equilibrium supercooled liquids.

In the last decade, the idea of freezing the positions of a set of particles has been investigated in various geometries~\cite{berthier2012static}, in addition to the closed cavity used to infer the point-to-set length scale. Freezing for instance the position of particles in a half space creates an infinite wall of frozen particles which acts as an interesting geometry to detect correlation length scales~\cite{scheidler2002growing,kob2012non,hocky2014crossovers}. Another example is when a finite fraction $c$ of particles chosen at random is frozen from an equilibrium configuration~\cite{kim2003effects}. In that case the system remains globally isotropic and spatially homogeneous but the frozen particles considerably reduce the size of the available configuration space. In the mean-field limit, it can be rigorously shown that this induces an entropy crisis with a nature similar to the temperature-driven Kauzmann transition~\cite{cammarota2012ideal}. As for the frozen cavity, the constraint imposed by this random pinning procedure makes it difficult to properly estimate thermodynamic properties of the remaining free particles, and parallel tempering has been used to study this situation. Evidence was provided that a sharp change is happening as $c$ is increased at constant temperature, which seems consistent with an incipient phase transition~\cite{kob2013probing}, accompanied by a drastic decrease of the configurational entropy~\cite{ozawa2015equilibrium}. In the future, lower temperatures should be studied and a rigorous finite-size scaling analysis should be conducted to fully establish that this situation truly corresponds to an equilibrium Kauzmann transition. 

\subsection{Rheology of amorphous solids}

\label{sec:rheology}

We now turn to the rheological properties of the glass state, which is a topic of obvious practical and experimental interest~\cite{meijer2005mechanical,schuh2007}. Computer simulations are well suited to analyse the glassy rheology of dense colloidal suspensions as the time scales that can be explored experimentally and numerically using conventional numerical methods coincide well. The analysis of steady state flow curves for materials undergoing large deformations represents an important research area~\cite{bonn}. 

By contrast, atomic and molecular glass-formers cannot be arbitrarily deformed since they break or fracture at large deformations. Therefore, one is led to analyse the elasticity of the glass in the linear response regime, the initial plasticity of the deformed material, possibly followed by the macroscopic failure which often takes the form of a macroscopic shear band where the plastic deformation is almost entirely localised. 

As usual, computer simulations are {\it a priori} severely limited in such an endeavour~\cite{rodney2011modeling}. A first issue is the typical timescale for the deformation of the material, which is obviously larger by many orders of magnitude in standard molecular dynamics than that which can commonly be reached in a real mechanical experiment. This problem was solved about twenty years ago by the introduction of a tool called athermal quasi-static deformation where incremental deformation steps are followed by a global energy minimisation~\cite{malandro1999relationships,utz2000atomistic,maloney2006amorphous}. In this approach, the effective rate of deformation is zero, and this particular timescale issue is completely solved, although thermal fluctuations are then neglected.

A more problematic issue is that computer simulations can only study the mechanical properties of configurations that are prepared numerically via some cooling protocol. Over the last two decades, computer simulations have therefore analysed the mechanical properties of molecular glasses quenched to the glass state with cooling rates that are about $10^8$ times faster than in conventional experiments, resulting in poorly annealed glassy states. Such systems can easily support large deformations and very much behave as soft colloidal glasses. That is, the yielding of the glass occurs as a smooth crossover to a flowing state as a function of deformation. This mode of yielding is typical of ductile materials, and has been carefully analysed in many simulation works~\cite{barrat2011heterogeneities}. 

\begin{figure}
\begin{center}
    \includegraphics[width=1.\columnwidth]{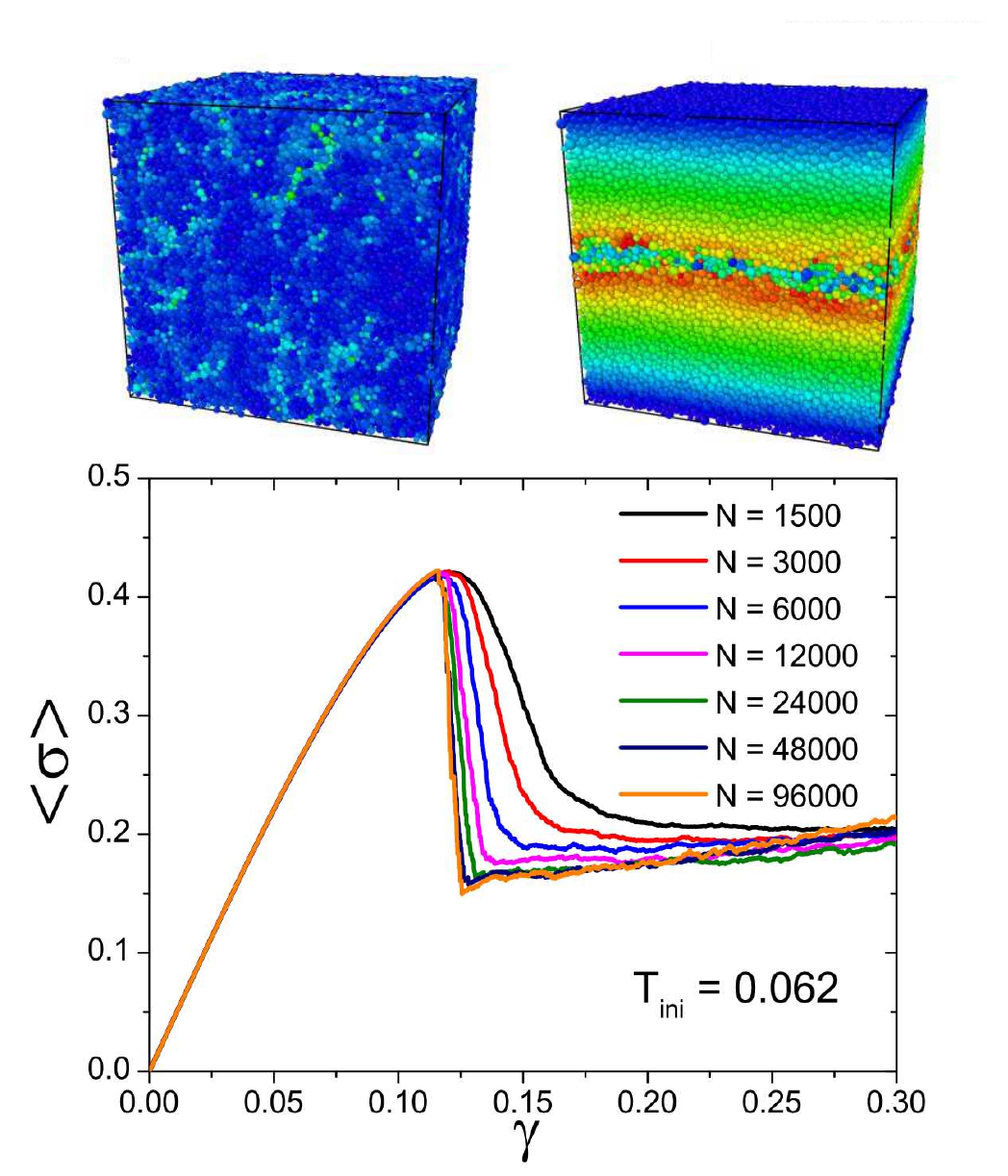}
\end{center}
    \caption{Brittle yielding of computer glasses via computer simulations. Top: ductile (left) and brittle (right) yielding. The color codes for the amount of plastic deformation in the deformed material at the yielding in transition in the case of a poorly-annealed ductile material and a very stable brittle material.
Bottom: The brittle yielding in a stable glass provokes a sharp and discontinuous macroscopic discontinuity is the stress-strain $\sigma(\gamma)$ relation in the thermodynamic limit. Adapted from Ref.~\cite{ozawa2018random}.}
    \label{fig:rheology}
\end{figure}

The advent of the SMC algorithm radically changed the situation, as the preparation of glassy configurations with effective cooling rates equivalent to the ones used for real molecular glasses now becomes possible. Employing the athermal quasi-static protocol for models of glass-formers similar to earlier work, it was first shown in 2018~\cite{ozawa2018random} that increasing the initial stability of the glass was sufficient to change the yielding behaviour from ductile as observed in earlier simulations to a brittle yielding accompanied by a macroscopic failure of the material, as seen in experiments. The tendency to localise the plastic deformation more strongly in space had been found to increase slowly as the preparation time of the system increases in previous work~\cite{shi2005strain}. However, for the very stable configurations analysed in Ref.~\cite{ozawa2018random}, it could be shown using a finite size scaling analysis that macroscopic failure happens suddenly in a single deformation step in a way that fluctuates less in larger systems, see Fig.~\ref{fig:rheology}. This is compatible with the view that yielding in very stable glassy states can be described as a kind of discontinuous non-equilibrium phase transition. This finding echoes both theoretical developments, where yielding is treated in the realm of statistical mechanics~\cite{rainone2015following,parisi2017shear}, and of course experimental results showing that real glasses break abruptly via brittle macroscopic failure.           

The initial work of Ozawa {\it et al.}\cite{ozawa2018random} paved the way for several research directions currently under intense scrutiny. In this initial study, the numerics suggested that the clear discontinuous yielding transition observed for very stable glasses becomes less discontinuous and slowly transforms into the smooth crossover observed for ductile materials. On the basis of these results complemented by the analytical solution of a mean-field model for yielding, it was suggested that the brittle-to-ductile yielding transition could itself be described by a second-order critical point with universal properties similar to the ones of a random field Ising model~\cite{ozawa2018random}. Further work in two dimensional glasses~\cite{ozawa2020role} seemed to confirm this picture, but concerns have later been raised about possible finite-size effects~\cite{barlow2020ductile,richard2021finite}. Future work employing off-lattice and more coarse-grained models~\cite{rossi2022finite}, together with theoretical developments, will hopefully clarify the nature of the brittle-to-ductile transition. 

A second line of research concerns variations of the geometry and timescale for the deformation. Recent studies have analysed in great depth the case of periodic deformation~\cite{yeh2020glass,bhaumik2021role} which represents an important class of mechanical tests in industrial applications. Also, the influence of a finite rate of deformation on brittle yielding was studied~\cite{singh2020brittle}. Finally, having at hand realistic glass samples, there is hope that a platform to understand in detail the microscopic nature events of the plastic events leading to stress relaxation in amorphous solids is now at hand~\cite{richard2020predicting,ozawa2022rare}. Additional studies should also address the effect of thermal fluctuations. 

\subsection{Vibrational and thermal properties of glasses} 

\label{sec:vibrational}

The solid state properties of glasses hold many mysteries which can now be convincingly addressed with the recent advent of the {\em in silico} preparation of well-equilibrated configurations. These features include the unusual properties of the vibrational density of states in a disordered solid~\cite{edigerangellnagel,gmmpv,schirmacher,elliott,sokolov,parshin,lerner2021low} as well as the putative role played by localized tunneling states on the energy landscape which become operative at cryogenic temperatures~\cite{ahv,phillips}.

\begin{figure}
\includegraphics[width=1.\columnwidth]{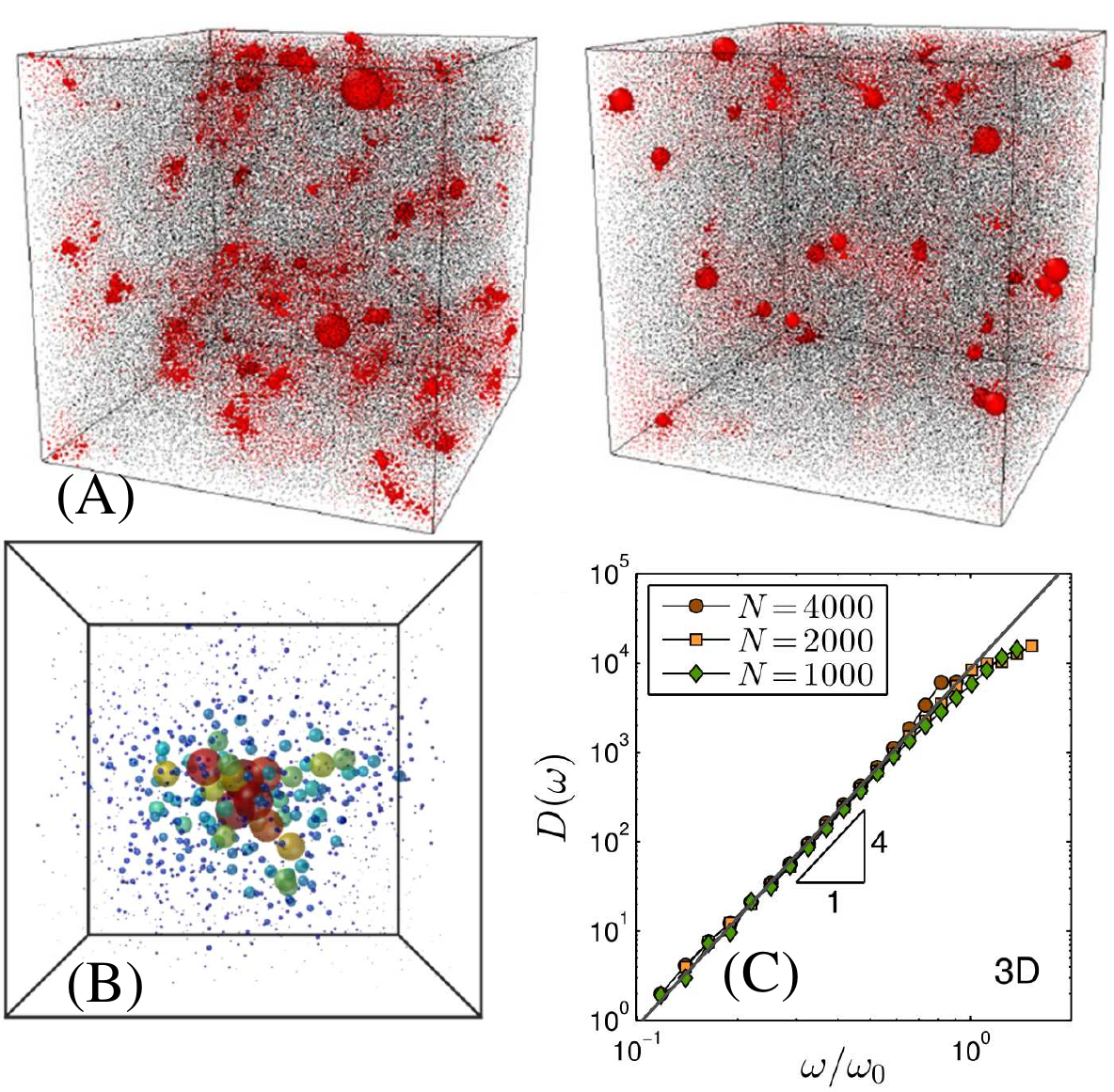}
    \caption{(A) The `softness' field obtained from summing the real-space character of low frequency modes for two different quench rates.  The configuration on the right (slower quench) illustrates the smaller number of quasi-localized modes with a higher degree of localization. Adapted from Ref.~\cite{wang2019low}. 
    (B) A rare, highly collective tunneling event in a simulated glass. Adapted from Ref.~\cite{khomenko2020depletion}
    (C) The density of states of vibrations in an amorphous solid illustrating the $\omega^{4}$ behavior at low frequency. Adapted from Ref.~\cite{kapteijns2018universal}.}
    \label{fig:tls}
\end{figure}

It has long been appreciated that configurational disorder induces stark qualitative changes in the behavior of the low-energy eigenstates of the vibrational Hessian matrix which quantifies quadratic fluctuations around a glass minimum~\cite{parshin,schober,ruocco}. In particular, in addition to the plane wave phonon modes expected from continuum elasticity, modes with localized behavior dubbed `quasi-localized' modes appear in the vibrational spectrum $D(\omega)$ (also called the density of states). The contributions of both extended and quasi-localized modes to $D(\omega)$ render its behavior distinct from the Debye spectrum expected for simple ordered solids.  A reasonable separation of the modes into extended and quasi-localized in the low frequency wing of $D(\omega)$ becomes apparent as the system size increases~\cite{wang2019low}.  A focus on the contribution of the extended modes to the spectrum reveals an excess peak (the `Boson peak' visible at some finite frequency when plotting the density of states rescaled by the Debye contribution) which then merges with the expected Debye behavior at much lower frequencies, namely $D_{\rm ex}(\omega) \sim \omega^{d-1}$, as $\omega \rightarrow 0$~\cite{kapteijns2018universal}.  At low-enough temperatures, the quasi-localized contribution to the spectrum surprisingly takes a different functional form $D_{\rm ql}(\omega) \sim \omega^{4}$, with the exponent independent of dimensionality~\cite{parshin, kapteijns2018universal, wang2019low}. These observations can be rationalized theoretically and are visible, to a lesser extent, in poorly annealed samples.  The advent of SMC has enabled the confirmation of these features in the well equilibrated samples of relevance to real experiments performed on amorphous solids.  Perhaps more importantly, the use of SMC has pinpointed which features evolve as a function of the annealing rate.  Specifically, the core of quasi-localized modes becomes more localized in better annealed configurations~\cite{lerner2018,wang2019low}. Concomitantly, the prefactor of the quartic law associated with the density of states of the quasi-localized modes decreases rapidly as the degree of annealing increases~\cite{wang2019low}.  The ultimate fate of these features as the quench rate continues to decrease remains an open question.  Several of these observations are illustrated in Fig.~\ref{fig:tls}.

At ultra-low temperatures near 1K, the thermodynamic properties of glasses markedly and quasi-universally deviate from those of crystals. In particular, the specific heat of the glass in this temperature range is much larger than in the parent crystal and shows a nearly linear, as opposed to cubic, temperature dependence~\cite{zeller}. Given the vibrational properties of amorphous solids discussed above, such behavior seems surprising since the contribution in excess to Debye due to quasi-localised modes does not account for the specific heat data. The thermodynamic properties of glasses in this regime have been explained by the highly successful but phenomenological two-level system theory of Anderson, Halperin and Varma~\cite{ahv}, and of Phillips~\cite{phillips}. This theory posits that local configurations in glasses can tunnel between two configurations, and the ensemble of these tunneling defects provides an excess set of excitations that explains the glass anomalies.

The first attempt to test this theory {\em in silico} and thus reveal the microscopic nature of local defect modes in glasses was made by Heuer and Silbey in 1993~\cite{heuersilbey}.  These authors investigated small ($N=150$) systems with a binary potential energy function that was created to mimic an amorphous mixture of nickel and phosphorous~\cite{weber2}.  Using standard molecular dynamics equilibration techniques, Heuer and Silbey located several hundred pairs of local minima separated by a single barrier.  However only a single such double-well potential had a tunneling splitting of the order of 1K. This difficulty necessitated the use of an indirect extrapolation technique to infer the distribution of tunneling levels in the range where the two-level system model is expected to be operative.  Regardless, the results of Heuer and Silbey conform closely to the predictions of the two-level system model. 

Over the intervening decades since this pioneering work, the approach taken by Heuer and Silbey has been extended and employed to study other glass-forming systems~\cite{daldoss1,daldoss2,heuer2,jug,rodney}.  Unfortunately limitations in algorithms and hardware have meant that computer simulations have produced an incomplete picture of low energy excitations in glasses.  In particular, in addition to small system sizes, the lack of an approach such as SMC has heretofore meant that unrealistically prepared {\em in silico} amorphous have been studied.

The advent of the SMC approach has provided an impetus to revisit the goals first laid out by Heuer and Silbey.  This challenge has been taken up by Khomenko {\em et al.}, who used SMC to investigate the local structure of the lowest energy states on the energy landscape of a polydisperse soft-sphere glass as a function of the degree of equlibration, ranging from systems which are poorly annealed to those that are as well equilibrated as laboratory ultrastable glasses~\cite{khomenko2020depletion,mocanu2022microscopic}. Khomenko {\em et al.} employ efficient protocols to locate connected minima and find the lowest energy pathways between them. Their approach has enabled the direct extraction of a large number of tunneling systems even under the most well-equilibrated conditions.  In turn, this has afforded a direct test of the two-level system model. Khomenko {\em et al.} find that the distribution of tunneling systems is in agreement with that proposed by Anderson, Halperin and Varma, and by Phillips.  In addition, they demonstrate that the density of tunneling systems decreases sharply as the stability of the glass configurations increases, as observed in many (but not all) experiments on ultrastable glasses~\cite{hellman,ramos,ramos2}.  Interestingly, Khomenko {\em et al.} show that while the vast majority of tunneling systems are associated with local defect-like motion, rare, highly collective tunneling motion occasionally occurs, especially in more poorly annealed systems, see Fig.~\ref{fig:tls}c.

Many unanswered questions remain with respect to our understanding of the non-phononic excitations discussed above.  Because SMC is limited in the range of potential energy surfaces for which it can provide great equilibration efficiency gains, the origin of the intriguing experimentally-observed quasi-universality in various ratios of material constants associated with two-level systems in starkly different glassy systems remains unexplored by computer simulation~\cite{leggett}.  The connection between two-level systems and quasi-localized modes, a connection that is central to the phenomenological soft-potential model~\cite{karpov,ilyin}, demands more attention despite some recent work along these lines~\cite{khomenko2021}.  A rigorous, multi-dimensional treatment of tunneling on the potential energy landscape has yet to be carried out~\cite{mills}. Lastly,  a precise, quantitative means of calculating the density of tunneling systems in simulated glasses is lacking due to the fact that protocols to search the energy landscape which are required to cull a statistically significant sample of two-level systems differ from the experimental quench pathway taking in the laboratory.  These and other issues and questions should be addressed in future work.

A surprising outcome of the large-$d$ statistical mechanics approach to the glass transition is the prediction that the glass phase itself is not unique, but can undergo a transition between two types of glassy states characterised by distinct physical properties~\cite{KPUZ13}. This Gardner transition had first been discovered in the context of mean-field spin glass models~\cite{Ga85} and its prediction in the context of structural glasses led to an intense research activity in recent years~\cite{berthier2019gardner}. From a computational viewpoint, SMC played a pivotal role in this endeavour because the predicted transition occurs when adiabatically following very stable glassy states that only SMC can achieve. Clear signs of a phase transition have been reported in three dimensional hard sphere glasses~\cite{BCJPSZ15,SZ18,JUZY18}, which becomes a strong crossover in two dimensional hard disks~\cite{LB18}. The phase behaviour of soft glasses is more subtle~\cite{scalliet2019nature}, and the Gardner transition does not seem to occur in more conventional glass-formers such as Lennard-Jones systems~\cite{SBZ17} showing that more work is needed to fully assess the problem of the Gardner transition in generic glass-formers.          

\section{Perspectives for future research}

\label{sec:perspectives}

From a fundamental perspective, the glass problem has gone through important transformations over the last few years, as theoretical, computational and experimental progress has been made, paving the way towards a better understanding of some key issues. We believe that this progress will help organise the field around well-posed questions that can directly be addressed analytically or numerically. We close this review by describing some research directions and goals where we feel important progress is likely to be made in the coming years. 

One nascent direction of research stems from the on-going revolution created by the systematic application of deep learning techniques to many areas across the physical sciences. The field of the glass transition has also been attacked by a variety of machine learning techniques to address various questions. One important line of research, which is somewhat peripheral to this article, is the use of machine learning tools to develop realistic interaction potentials between glass-forming atoms and molecules which have an {\em ab-initio} level of accuracy (see, for example,~\cite{deringer2021origins}), thus making important progress towards the development of better models in the first category described in Sec.~\ref{sec:basic}. A second line of research involving machine learning pertains to the development of methods to detect, in an unsupervised manner, the existence of important structural properties in glassy configurations that may otherwise appear devoid of any structural heterogeneity to the naked eye~\cite{paret2020assessing,boattini2020autonomously}. The goal here is to use deep learning techniques to automatically detect, with no {\em a priori} bias, the geometric motifs that may be relevant to understanding the thermodynamic evolution of glass-formers, the hope being that machine learning can outperform the existing attempts to detect relevant structural motifs in supercooled liquids that are based on physical intuition~\cite{coslovich2007understanding,malins2013identification,tong2018revealing}. Finally, a related but somewhat distinct investigative thrust is the application of machine learning techniques to probe the fundamental question of how the structure of a supercooled liquid encodes the heterogeneous slow dynamics arising from a given initial configuration. Simple metrics, such as the softness field which relies on training local dynamics to local structural pair correlation functions~\cite{schoenholz2016structural}, and more complex approaches such as the use of graph neural networks~\cite{bapst2020unveiling} to predict the structural propensity for dynamics on different timescales have been proposed. Current efforts attempt to develop the best architecture to improve the quality and the simplicity of the predictions, while at the same time improving on the list of structural indicators used as inputs~\cite{alkemade2022comparing}. The overall goal would be both to make extremely accurate predictions in order to draw some physical conclusions from them, and to extract from the learned models what are the structural descriptors which correlate best with the long-time dynamics. Given the pace at which machine learning techniques are propagating throughout many computational areas, we expect to see much further activity in this area. These approaches have, in particular, great potential to allow for a detailed mechanistic understanding of how the structure (e.g., local packing motifs), which appears remarkably similar to that of the high-temperature liquid state that it was cooled from, encodes the dramatically heterogeneous dynamics that is the hallmark of supercooled liquids. Similar questions are also being asked in the context of deformed glasses where plasticity is typically also very heterogeneous in space and time~\cite{cubuk2015identifying,richard2020predicting}. 

There are of course a number of open questions that are awaiting the development of better simulation techniques to be attacked with greater vigor. Techniques such as SMC work remarkably well for systems that mimic metallic glass-formers, but the simulation of glasses composed of the more complex family of molecular glass-forming liquids are currently out of reach via any Monte Carlo approach. Can generalized versions of cluster Monte Carlo methods be invented to simulate such systems? The ability to do so would help answer the important question of how universal the underlying microscopic dynamical motifs are in a diverse class of glass-forming systems. The many new results concerning the physical behaviour of simple point-particle models near the experimental glass transitions have not yet been confirmed in more complicated but experimentally typical models. Although there is hope and theoretical reasons why a fair degree of universality will eventually be found, it is mandatory to develop the numerical tools that will confirm this hypothesis. This would then allow researchers to close the remaining gap between simulations and experiments. We believe that extended versions of the SMC could first be attempted in the simplest known models of molecular glasses, such as systems comprised of coarse-grained molecules or short polymeric chains.  

The recent progress using computer simulations to directly determine the thermodynamic fluctuations and extended phase diagrams of supercooled liquids near the glass transition have demonstrated that the framework of RFOT theory, which stems from the firm basis of the mean-field theory of the glass transition, appears to correctly describe the thermodynamics of two- and three-dimensional glass-formers. However, the strong hints of an underlying Kauzmann transition remain subject to a temperature extrapolation, because equilibrating to temperatures that are close enough to $T_K$ remains impossible. It could be that the SMC is too primitive an algorithm to approach $T_K$ because it represents, after all, only a simple variation upon canonical Monte Carlo simulations for fluids. We may surmise that the development of smarter Monte Carlo algorithms which possibly displace several particles at once, perhaps in conjunction with parallel tempering, would be needed to approach the Kauzmann transition closely enough to directly observe the entropy crisis and firmly test the validity of RFOT theory.   

Connected to this question, there is a pressing need for the development of approaches that may, perhaps in a coarse-grained manner, enable the simulation of the long-time dynamics of supercooled liquids close to the glass transition. Over the last decade great progress has been made in the creation of dynamical strategies~\cite{eaves,charb,hoy2022efficient,berthier2019bypassing}, correlation functions~\cite{berthier2007-1, berthier2007-2} and tools~\cite{harowell1,harowell2} to assess and describe crucial aspects of glass formation, such as dynamical heterogeneity and growing dynamical length scales.  Some of these tools are expensive to simulate via molecular dynamics even at relatively high temperatures.  For example, the quantity $S_{4}({\bf q},t)$, which is a four-point function from which growing dynamical length scales can be extracted~\cite{berthier2007-1,berthier2007-2}, requires not only a large degree of ensemble averaging but also large system sizes to extract accurate dynamical exponents~\cite{karmakar}. The advent of techniques such as SMC enable to creation of glassy samples that are annealed in a realistic manner beyond the capability of local Monte Carlo or molecular dynamics. However, simulating the direct long-time dynamics from such configurations remains a difficult task for highly annealed systems because the dynamics themselves span many decades in time. The use of advanced Monte Carlo methods to create initial conditions for simulations on dedicated specialized hardware such as Anton can only partially overcome this problem~\cite{shaw}. While SMC has helped enlarge the time window that can actually be analysed numerically~\cite{berthier2021self,guiselin2022microscopic,scalliet2022thirty}, future effort must be focused on this most crucial of questions. In our view, the dynamics of deeply supercooled liquids in the temperature regime pertinent to experimental work, remains a largely unexplored territory. Although measured thermodynamic fluctuations tend to agree with the RFOT theory description, this gives no guarantee that relaxation dynamics directly follows from the thermodynamics in the manner envisioned by theory~\cite{RFOT,bouchaudbiroli}, or, more broadly, that it is controlled by features encoded in the potential energy landscape~\cite{middleton2001energy,heuer2008exploring}. In particular there are several indications that dynamics may be controlled by a small population of localised defects~\cite{isobe2016applicability,keys2011excitations} and that purely dynamic relaxation channels that are not described by thermodynamic quantities contribute to the long-time dynamics such as dynamic facilitation~\cite{bergroth2005examination,vogel2004spatially,keys2011excitations,scalliet2022thirty}. We expect that the development of novel algorithms together with improvement in molecular dynamics implementations will provide the tools needed to numerically study the long time dynamics of glassy liquids near the experimental glass transition.  

\begin{acknowledgments}
This work was supported by a grant from the Simons Foundation (\#454933, LB, \#454951 DRR) and by a Visiting Professorship from the Leverhulme Trust (VP1-2019-029, LB). The authors thank all of the members of the Simons Foundation Collaboration on "Cracking the Glass Problem" for six+ years of stimulating discussions.
\end{acknowledgments}

\bibliography{main.bib}

\end{document}